\documentclass[12pt,a4paper]{article}
\usepackage[utf8]{inputenc}
\usepackage[english]{babel}
\usepackage{amsmath}
\usepackage{amsfonts}
\usepackage{amssymb}
\usepackage{graphicx}
\usepackage{authblk}

\author[$\dagger$]{Francisco Gonzalez Montoya}
\author[$\star$]{Christof Jung}

\affil[$\dagger$]{\footnotesize School of Mathematics, University of Bristol, Bristol, BS8 1UG, United Kingdom }

\affil[$\star$]{\footnotesize Instituto de Ciencias Físicas, Universidad Nacional Autónoma de México,  Av. Universidad s/n, Cuernavaca 62251, México }

\date{ }

\title{ The numerical search for the internal dynamics of NHIMs and their 
pictorial representation}

\begin{document}
\maketitle

\begin{abstract}

The topic of this article is the numerical search of codimension 2 Normally Hyperbolic Invariant Manifolds (NHIM) in Hamiltonian systems with 3 degrees of freedom and their internal dynamics. We point out relations between different strategies to find such surfaces numerically. We can start from index-1 saddles of the effective potential or from a partially integrable case and follow the NHIM along some curve in the parameter space of the system. Or, we can look for the stable and unstable manifolds of such surfaces by an appropriate indicator method. We show numerical examples for an electron moving in a perturbed magnetic dipole field.

\end{abstract}

\section{Introduction}

The study of phase space structures is a fundamental problem in dynamical systems. It is essential to understand the behaviour of the trajectories from a theoretical and practical perspective. The traditional tools to visualize the phase space structures like the projection of the trajectories onto a plane or Poincar\'e maps are important to understand diverse properties of the phase space. However, the study of the phase space structures of multidimensional systems remains an open and challenging problem. New tools have been developed to study the phase space structures of multidimensional systems like Fast Lyapunov Exponents (FLE) \cite{leg},  Mean Exponential Growth Factor of Nearby Orbits (MEGNO) \cite{cin}, Smaller  Alignment Indices (SALI), Generalized Alignment Indices (GALI) \cite{sko}, and Lagrangian Descriptors (LDs) \cite{mad,l}. The phase space structure indicators are scalar fields constructed with the trajectories of the system. The differences in the values of the scalar fields give us information of the phase spaces objects that intersect the set of trajectories considered. One of the most intuitive kind of indicators are the Lagrangian descriptors based on trajectories' arc length; some recent examples are presented in the references \cite{a,gw1,gw2,gw3,f1,f2}. The differences in the arc length of the trajectories with nearby initial conditions give us information about the phase space around those initial points. In this work, we use a quantity related to the arc length, the delay time. The delay time is a natural indicator to study the dynamics of open systems.

The topic of this article is the numerical search for normally hyperbolic invariant manifolds, usual abbreviation NHIMs, their internal dynamics and their pictorial representation. We concentrate on Hamiltonian systems with three degrees of freedom, abbreviation 3-dof, and work to a large extent with the Poincar\'e map of the system, which has a 4-dimensional domain for a given value of the energy. Dynamical systems have some skeleton that directs the global dynamics to a large extent. In a 2-dimensional map, the most important bones of the skeleton are the fixed points. The stable ones are centres of islands of regular motion. And the unstable ones are the organizing centres of chaotic regions. These unstable fixed points have stable and unstable manifolds, which build up horseshoe structures in the  Poincar\'e map. For the importance of horseshoes in dynamical systems, see section 3.2 in \cite{hu1} and for a pictorial explanation of them, see chapters 13 and 14 in \cite{hu2}. Note that the fixed points are of codimension 2 in the map domain. Accordingly, the stable and unstable manifolds of hyperbolic fixed points are of codimension 1. Therefore, they form local divisions between regions of different behaviour, and they can channel and guide the dynamics in the corresponding chaotic region of the domain of the map.

When we go over to 4-dimensional maps, then a natural question arises: which subsets, in this case, take over the role that the hyperbolic fixed points play in 2-dimensional maps? A first but wrong guess might be that, again, the completely hyperbolic fixed points play this key role. However, in 4 dimensions, the stable and unstable manifolds of hyperbolic-hyperbolic or also of complex unstable fixed points have dimension 2. Therefore, they are of codimension 2 in the 4-dimensional domain of the map, and accordingly, they do not form local divisions. To provide local divisions, we need surfaces of codimension 1. And such dividing surfaces of codimension 1 can be the stable and unstable manifolds of invariant subsets of dimension 2 ( and thereby also codimension 2 ), which are hyperbolic in normal directions. That is, they are normally hyperbolic invariant surfaces of codimension 2. These surfaces are the topic of the present article. The principal tasks addressed in this article are the following: Find 2-dimensional normally hyperbolic invariant manifolds in a 4-dimensional Hamiltonian map, find the internal dynamics of these NHIMs and present it graphically.

Of course, it is not easy to find such surfaces as long as we do not have any reasonable idea where to start the search and what shape to search for. We need some clue in which parts of the phase space we begin our search and what approximately should be the shape of this surface. In section 2, we discuss two particular situations where we know beforehand a rather good approximation for the NHIM. Then the initial step is a numerical construction of the  NHIM for these simple particular starting cases. Next, we need the idea of how to follow the NHIM numerically under perturbations of the simple initial cases. This can be done by following the stable manifold of the NHIM. The existence of the NHIMs and their stable manifolds under sufficiently small perturbations is guaranteed by the persistence theorem, see \cite{fen, wig, CK}. In section 3, we demonstrate numerically how a plot of the delay time over some surface of the domain of the map identifies these stable manifolds. In section 4, we show how a further selection of only the local segments of the stable manifold allows to identify the internal dynamics of the NHIMs and their graphical representation. As an example of demonstration, we use the motion of an electron in the field of a magnetic dipole perturbed by a magnetic quadrupole. In section 5, we give final remarks and conclusions.

\section{The particular parameter case to start from}

The problem is to search for a 2-dimensional NHIM surface in a 4-dimensional Poincar\'e map. This may be not easy if we do not know in which region of the domain of the map to start the search. Very helpful is some information about the location of the NHIM for particular parameter cases. Then, we can start from these particular cases and follow the NHIM under general changes of the system parameters. During changes of the parameters, we always have the idea in mind that thanks to the persistence theorem, small changes of the parameters cause small changes of the shape and the location of the NHIM only. The persistence theorem holds as long as the tangential instability on the NHIM is smaller than its normal instability.

So far, we are aware of 2 very different possibilities to start at particular parameter values for which we know approximately the location of the NHIM. The first one is a start at an energy value just a little bit above the energy of an index-1 saddle of the effective potential of the system. The second one is a start at a parameter case that is partially integrable. In the following subsections, we first present these two possibilities and then give some comments on the connection of the second possibility with the first one.

\subsection{Index-1 saddle of the effective potential}

Assume a Hamiltonian 3-dof system with position coordinates $(q_1, q_2, q_3)$ and canonically conjugate momentum coordinates $(p_1, p_2, p_3)$. In the following, we just write $q$ or $p$ when we mean the collection of the 3 components. The Hamiltonian function is given as $H(p,q)$. The effective potential of the system as a function of the position coordinates is given by 

\begin{equation}
V_{eff} (q) = \min_p H(p,q).
\end{equation}

That is, we take the minimum of the Hamiltonian over the momentum at a fixed position, see also \cite{sm, ma}. Now assume that $V_{eff}(q)$ has a nondegenerate index-1 saddle point at $q_s$ and that $V_{eff}(q_s) = E_s$. For total energies $E$ just a little above $E_s$, it is a good approximation to expand $V_{eff}(q)$ in a power series around $q_s$ and to truncate the series after the quadratic terms. In this approximation, we can calculate analytical solutions of the equations of motion. For an index-1 saddle of a 3-dof system, we obtain two very special solutions over this saddle; the Lyapunov orbits \cite{ly}. They are periodic orbits that oscillate over the saddle and are dynamically unstable in the unstable direction of the potential. Furthermore, there is a 2-dimensional continuum of quasiperiodic trajectories over the saddle. We can imagine that over a saddle point of the potential of a 3-dof system, there exist 3 normal modes. For an index-1 saddle, one of them belongs to the unstable direction of the potential and leads to trajectories that run away from the saddle exponentially. The other two normal modes are oscillatory, and the pure form of an oscillatory normal mode is the corresponding Lyapunov orbit.

We can understand the existence of the continuum of localized trajectories as follows. Imagine we treat the energy value $E=E_s + \Delta E$, and we want to excite the stable oscillatory modes over the saddle only. Then we have a 1-dimensional continuum of possibilities to distribute the energy $\Delta E$ between the two oscillatory modes. And in addition, we have a 1-dimensional continuum of possibilities for the initial relative phase shift between the two oscillatory modes. In total, we obtain a 2-dimensional continuum of trajectories localized over the saddle and consisting of quasiperiodic superpositions of the two localized normal modes. This set of localized trajectories forms a 3-dimensional surface in the 5-dimensional energy shell. It is the NHIM in the energy shell. In quadratic approximation, it is a 3-dimensional ellipsoid. This surface has two important properties. First, it is invariant under the dynamics. An initial point on this surface leads to a trajectory that stays on this surface for all times. And second, this surface is dynamically unstable (hyperbolic) in directions normal to the surface, i.e. in directions of the third, unstable normal mode of the potential saddle. These two properties of the surface explain the origin of the name ``Normally Hyperbolic Invariant Manifold''.

In many cases, it is more convenient to handle Poincar\'e maps instead of flows. Without loss of generality, we assume that the intersection condition for the map is $q_3=0$. In general, intersections with one particular orientation have to be chosen, only in the case of a discrete symmetry $q_3 \rightarrow -q_3$ the orientation is irrelevant, and the intersections of both orientations can be used. For a 3-dof system at fixed energy, the domain of the map has dimension 4. In the map, the NHIM is represented by a 2-dimensional surface. In section 4 of \cite{jz1}, the analytical construction of the NHIM in quadratic approximation has been presented in all details for an example from celestial mechanics.

\subsection{Stack construction in the partially integrable case}

Other situations where we know the approximate position of the important NHIM beforehand are the cases where the system we are interested in is close to a partially integrable case. Then we treat first the partially integrable limiting case, and later we switch on the nonintegrable perturbation and follow the NHIM along the perturbation. In this subsection, we concentrate on the partially integrable case. That is, we consider the case where there is a second independent conserved quantity besides the energy, but there is no third independent conserved quantity. We take as an example of demonstration the motion of an electron in a magnetic field where the partially integrable case is the pure dipole field and the perturbation is the addition of higher multipole contributions which destroy the rotational symmetry of the pure dipole case, see also \cite{gj1, gj2}. 

Assume that the given 3-dof Hamiltonian $H(q_1, q_2, q_3, p_1, p_2, p_3)$ does not depend on $q_2$. Accordingly $p_2$ is a conserved quantity and $q_2$ is a cyclic position variable. As the intersection condition for the Poincar\'e map $M$ we still use $q_3 = 0$ with appropriate orientation conditions as explained at the end of the last subsection. The domain $D$ of $M$ has the coordinates $q_1, q_2, p_1, p_2$.

The whole phase space and the domain $D$ have a foliation into invariant leaves of constant $p_2$. The reduced dynamics inside of each leave is generated by the 2-dof Hamiltonian $H_{2-dof}(q_1, q_3, p_1, p_3, c)$ where $c$ is a fixed value of $p_2$ and thereby $p_2$ acts as a parameter in this reduced dynamics. To this reduced dynamics belongs a reduced 2-dimensional Poincar\'e map $M_{2-dof}(c)$ having the $q_1$-$p_1$ plane as domain. The intersection condition of $M_{2-dof}(c)$ is the same as the one for the full map $M$.

Now, the full-dimensional map $M$ can be reconstructed from the reduced map by the following stack construction. First we construct the continuum of 2-dimensional Poincar\'e maps $M_{2-dof}(p_2)$. Assume that this dimension reduced map has a hyperbolic fixed point $X(p_2)$ in the interval $p_2 \in [a, b]$. Next, we pile up the 2-dimensional maps where we use $p_2$ as stack parameter. The stack of the fixed points $X(p_2)$ is given by a curve $C$ in this stack which projects 1:1 onto the $p_2$ axis.

Finally, we form the Cartesian product of the stack with the manifold $A$ of the possible values of $q_2$. Depending on whether the second degree of freedom is open or closed, this manifold $A$ can be either a line, or a half-line, or a finite interval, or a circle. The Cartesian product of $C$ with $A$ results in a 2 dimensional surface $S$ lying in $D$. By construction, this surface is invariant, and because it originates from a hyperbolic fixed point in the $q_1$-$p_1$ plane, also $S$ is hyperbolic in the $q_1, p_1$ directions, i.e. it is hyperbolic in its normal directions. Then $S$ is the NHIM for the Poincar\'e map that we have been looking for.

Next, we demonstrate the procedure for the example of the motion of a charged particle in a magnetic dipole field. The partial integrability becomes evident due to the symmetry of the magnetic field with respect to rotations around the axis defined by the magnetic moment of the dipole.

For this case, the Hamiltonian in cylindrical coordinates is given as
\begin{equation}
H = \frac{1}{2m} \left( p_\rho^2 + p_z^2 + \left( \frac{L_z}{\rho} - \frac{e}{c} A_{\phi}(\rho,z) \right)^2 \right). 
\end{equation}
% Gl.2

\noindent The $\phi$ component of the magnetic dipole potential is given as

\begin{equation}
A_{\phi} = A_0 \frac{ \rho} { (\rho^2 + z^2)^{3/2}},
\end{equation}
% Gl.3

\noindent where $A_0$ is the dipole moment, $e$ is the charge of the electron and $m$ its mass, and $c$ is the speed of light. As explained in \cite{sc}, all these 4 constants can be scaled away. Therefore, we will set them to the value 1. The other components of the magnetic potential are zero for a pure dipole aligned in the $z$-direction.

The angle $\phi$ does not appear in $H$. Therefore, $L_z$ is a conserved quantity that we can handle as if it would be a parameter. We can separate away the $\phi$ degree of freedom and are left with a 2-dof system whose Hamiltonian is still given by Eq. 2. The corresponding effective potential is given by setting $p_{\rho} = 0$ and $p_z=0$ in Eq. 2.

\begin{equation}
V_{L_z}(\rho,z) = \frac{1}{2}\left(\frac{L_z}{\rho} - \frac{\rho}{(\rho^2 + z^2)^{3/2}} \right)^2.
\end{equation}
% Gl.4

\begin{figure}[htbp]
\begin{center}

\includegraphics[scale=0.2]{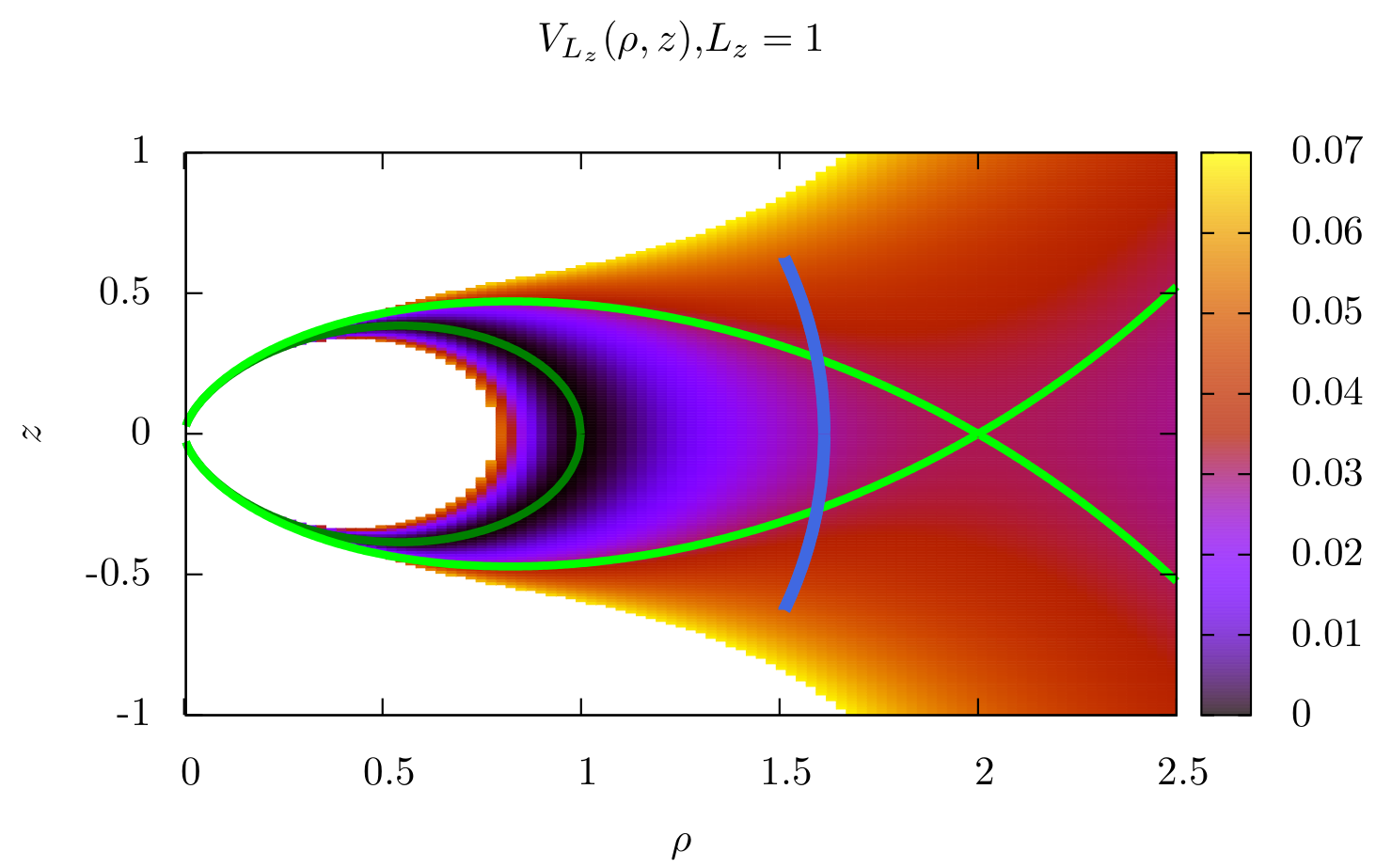}

\caption{ The effective potential $V_{L_z}(\rho,z)$ for $L_z=1$ is plotted in colour scale. The light green curve is the energy level set of the saddle energy $E_s = L^4/32 = 0.03125$. This curve has a selfintersection at the saddle point $q_S=(\rho,z)=(2,0)$. The dark green curve is the minimum of the potential, i.e. $V_{L_z}(\rho,z)=0$. The blue curve is the Lyapunov orbit $\gamma_{L_z}$ over the saddle for $E=0.05$. This unstable periodic orbit corresponds to the hyperbolic fixed point $X_{L_z}$ in the Poincar\'e map with intersection condition $z=0$ and $E=0.05$.}\label{fig:potential}
\end{center}
\end{figure}

The Fig. \ref{fig:potential} shows a plot of the potential energy $V_{L_z}(\rho,z)$ in colour scale for $L_z=1$. This potential has a saddle point $q_S$ at $\rho=2/L_z, z=0$ with a saddle energy $E_s = L^4 / 32$. Over the potential saddle we have an unstable periodic orbit $\gamma_{L_z}$, the Lyapunov orbit, which for a given total energy $E$ exists in the parameter interval $L_z \in [L_{min}, L_{max}]$ where $L_{max} = (32 E)^{1/4}$ and $L_{min}$ can only be determined numerically as $L_{min} \approx \, 0.788 L_{max}$. At $L_{max}$, the Lyapunov orbit  $\gamma_{L_z}$ disappears in the energetic boundary, and at $L_{min}$ it disappears in a saddle-centre bifurcation colliding with an elliptic periodic orbit. Many more details on these events are explained in \cite{js, gj1, gj2}. The Lyapunov orbit corresponds mainly to an oscillatory motion in the $z$-direction, and in the 2-dimensional Poincar\'e map for the reduced system, it appears as a hyperbolic fixed point. For the Poincar\'e map, we choose the intersection condition $z=0$ where the orientation is irrelevant because of the reflection symmetry in $z$.

Now we look at the energy shell of the full 3-dof system. This 5-dimensional constant energy manifold has a natural invariant foliation into the leaves of constant $L_z$. We can imagine a stack construction where $L_z$ is the stack parameter. Consequently, the domain of the full Poincar\'e map has this stack structure. The stack of the hyperbolic fixed points $X_{L_z}$ of the reduced map gives in a first step a curve that extends over the interval $[ L_{min}, L_{max} ]$. And in a second step, the inclusion of the cyclic angle $\phi$ gives a Cartesian product of this curve segment with a circle $S^1$. Thereby the surface in the 4-dimensional domain of the full map is a 2-dimensional cylinder segment. This cylinder segment is the NHIM $\mathcal{M}$ in the Poincar\'e map for the partially integrable case and is given by 

\begin{equation}
\mathcal{M} = \bigcup_{L_z} X_{L_z} \times S^1.
\end{equation}

\begin{figure}[htbp]
\begin{center}

\includegraphics[scale=0.2]{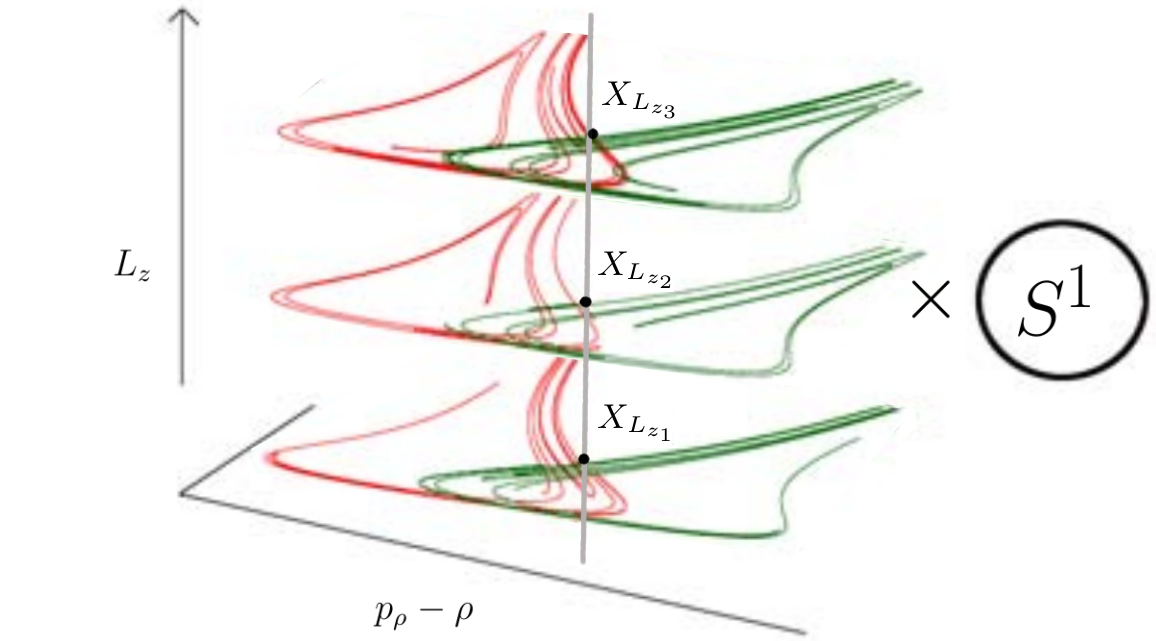}

\caption{ Poincar\'e map of the 3-dof symmetric system. 
The Poincar\'e map is constructed as a union of the Poincar\'e maps of the 2-dof systems parametrized by the conserved momentum $L_z$ and taking the cartesian product with the unitary circle $S^1$ that represents their conjugate angle $\phi$. The grey line is formed by the hyperbolic fixed points $X_{L_z}$. The stable and unstable manifolds $W^{s/u}(\mathcal{M})$, represented as green and red and lines, respectively, form spirals when $L_z$ grows. }\label{fig:stack}
\end{center}
\end{figure}

Naturally, $\mathcal{M}$ has stable and unstable manifolds that we can construct in an analogous way, see Fig. \ref{fig:stack}. Let us consider the union of all the stable and unstable manifolds of the hyperbolic fixed points $X_{L_z}$ parametrized by the conserved quantity $L_z$ and take the cartesian product with the circle $S^1$ that represents their conjugate angle $\phi$,

\begin{equation}
W^{s/u}(\mathcal{M}) = \bigcup_{L_z} W^{s/u} (X_{L_z}) \times S^1.
\end{equation}

The NHIM $\mathcal{M}$ for the Poincar\'e map corresponds to an unstable invariant surface for the flow in the constant energy manifold. It is the starting point to follow the NHIM surface through the perturbation of the partial integrability. For a lot more details, see \cite{gj2}.

\subsection{Effective potential for the partially integrable case}

In the last two subsections, we have presented two possibilities to find NHIMs for particular parameter cases, which we can use as a starting point to investigate the development scenario of a NHIM. The first possibility works for energies just a little above an index-1 saddle of the effective potential, and the second possibility is appropriate for systems close to a partially integrable case. Now, the reader may ask the following: Can we also apply the first method to a partially integrable case? Does it lead to the same NHIM as the second method? Which one of the two methods is easier to apply?

If we try to apply the first method to the Hamiltonian $H$ described in the second paragraph of subsection 2.2, then we run immediately into the following problem. This Hamiltonian does not really depend on the position coordinate $q_2$. Thus the minimalization of this Hamiltonian with respect to the momentum also leads to an effective potential that does not depend on the position coordinate $q_2$, i.e. the application of Eq.1 does not lead to a potential having any nondegenerate extremal point. We can avoid this difficulty by the application of a canonical transformation that interchanges the roles of $q_2$ and $p_2$, i.e. we apply the canonical transformation $p_2 \to q_2$ and $q_2 \to -p_2$. Then the minimalization with respect to the new momentum leads to an effective potential that depends explicitly on all three new position coordinates. Of course, it coincides with the effective potential that we obtain when we treat $p_2$ as a parameter and minimalize the original Hamiltonian with respect to $p_1$ and $p_3$ only. Therefore it is obvious that both methods lead to the same effective potential and, in the end, to the same NHIM.

Motivated by these questions and problems, we try to apply the first method to the example of the magnetic dipole. We will see that this particular example shows interesting additional difficulties. The Hamiltonian for this system in Cartesian coordinates is

\begin{equation}
H(q,p) = \sum_{j=1}^3 \frac{1}{2m}\left(p_j - \frac{e}{c} A_j(q) \right)^2.
\end{equation}
% Gl.5

For any value of $q$ we obtain the minimal possible value of $H$ by setting $p_j = e A_j(q) / c$ and this minimal value is always 0. That is $V_{L_z} = 0$ for a charged particle in a magnetic field. Of course, as anticipated, this potential does not have any nondegenerate index-1 saddle points, and the method from subsection 1 can not be applied in its original form. An effective potential identically zero for a charged particle in any magnetic field is consistent with the properties of the magnetic forces. Namely, magnetic forces are always perpendicular to the velocity of the particle. Therefore the speed of the particle is constant; only the direction of the velocity changes. As a consequence, the kinetic energy of the particle is conserved, and this means that the potential energy is a constant, which we always can take equal to zero.  

Now, we are led naturally to the following question: How can we obtain the effective potential from Eq. 4 by applying the rule from Eq. 1 to the dipole case? First, from Eq. 4, it becomes clear that we should use cylindrical coordinates, which are adapted to the symmetry of the dipole field. Second, in Eq. 1, we keep the position coordinates fixed, and in Eq. 4, the angular momentum $L_z$ acts as a parameter, i.e. it is a quantity that is kept fixed. And the conjugate angle $\phi$ is not the appropriate candidate to be fixed. This suggests the following idea, already recognized for the general case: Apply a canonical transformation that interchanges the roles of $\phi$ and $L_z$, i.e. we take $q_2 = L_z$ and $p_2 = - \phi$. Now, an application of the rule from Eq. 1 immediately leads to the effective potential form Eq. 4. However, this potential does not have any nondegenerate index-1 saddle in its 3-dimensional domain. Therefore, the rule from subsection 1 to construct an approximation for the NHIM  $\mathcal{M}$ still does not work.

Here the following idea helps: The force in $\rho$ direction is zero at $\rho_s = 2/L_z$. Accordingly for $L_z= L_{max}$ there exists the periodic orbit with $\rho = \rho_s$ and $z, p_z,$ and $p_{\rho}$ all identically zero and a constant angular velocity $L_z^3/8$ in $\phi$ direction. In the map, it is an invariant circle going around in $\phi$ direction at constant values of the other 3 coordinates $(\rho, p_{\rho}, L_z)$ of the domain of the map. This circle is the upper boundary of the NHIM $\mathcal{M}$ in the map, and it is dynamically unstable in the $\rho$ direction. For any $L_z$ value a little smaller than $L_{max}$ there must be some value of $\rho$ close to $\rho_s$ that belongs to an invariant circle of the map for this value of $L_z$, etc. Thereby starting from $L_{max}$, we can work our way down in $L_z$ numerically in small steps until we approach the so far unknown $L_{min}$. We notice the approach to $L_{min}$ by monitoring the normal instability as a function of $L_z$. When $L_z$ approaches $L_{min}$, then the normal eigenvalue approaches 1 from above. Numerically we simply check the loss of instability in the $\rho$ direction. So we obtain the cylinder segment of the NHIM $\mathcal{M}$ numerically in the Poincar\'e map.

The example of the dipole suggests the following. The application of the first method to a partially integrable case can lead to tricky problems. Index-1 saddles of the effective potential for the reduced system are not necessarily related to index-1 saddles of the effective potential for the full-dimensional system. If we can make a dimension reduction of the system, then it leads to a simpler system. This is consistent with the general recommendation to apply dimensional reductions whenever this is possible. The second method is easier to understand and to perform when we have some partially integrable system to start from. In previous numerical investigations of the development scenarios of NHIMs in 3-dof systems, we used successfully and without running into unexpected difficulties, the second method for the magnetic dipole problem \cite{gj2} and for a model map also having a rotationally symmetric limit case to start with \cite{dgj}. In contrast, we used the first method for several examples from celestial mechanics \cite{gj1,zj2,zj3,zj4,c21} and also for the hydrogen atom in an external rotating field \cite{jw}, example systems far away from any partial integrability.

The concluding recommendation is: Whenever we have a nearby partially integrable system, then we use the second method. The first method is the only choice and works well if the system is far away from partial integrability. If neither of these two strategies is appropriate for a given system, then the application of more general indicator methods might be the natural approach. The application of indicators is the theme of the next section.

\section{The delay time as an indicator of stable manifolds of invariant subsets for the symmetric and perturbed cases}

The motivation to introduce a delay time to study the phase space comes from the problem of chaotic scattering. The reference \cite{ssj1} is a review on chaotic scattering, and some recent examples of chaotic scattering can be found in \cite{bssj1, fssj, nzssj,k,dgjt,gbj}. When an incoming scattering trajectory runs exactly on the stable manifold of a localized subset sitting in the interaction region, then the trajectory will converge to this localized subset and gets stuck in the interaction region forever. When the trajectory does not start exactly on the stable manifold but close to it, then it comes close to the localized subset, stays some time in its neighbourhood before it leaves the interaction region again, and returns to the asymptotic region. The trajectory has a time delay in the interaction region. Thereby in the set of all initial asymptotes, the time delay in the interaction region is an indicator for the stable manifolds of localized subsets.

Let us assume we prepare the initial conditions of the trajectory on an incoming asymptote at time $t_i$, at a point $r_i$ and with velocity $v_i$ running in the direction towards the interaction region. Take the point $r_0$ as some reference point in the interaction region. Then the initial distance from the interaction region is $d_i = |r_i - r_0|$ and the particle needs the time $d_i / v_i$ to reach the interaction region. We measure the outgoing asymptote at time $t_f$, at a point $r_f$ leaving the interaction region with velocity $v_f$. Accordingly, the particle spent the time $d_f / v_f$ along the outgoing asymptote. The time delay in the interaction region is the elapsed total time minus the time spent along the incoming and outgoing asymptotes. In total, we find the time delay $\Delta t$ in the interaction region as

\begin{equation}
\Delta t = t_f - t_i - \frac{d_f}{v_f} - \frac{d_i}{v_i}.
\end{equation}

Note that according to this definition, the time delay is independent of the exact point along the asymptotes where we prepare and measure the asymptotic motion.

Now, let us construct an indicator in analogy to these ideas coming from
chaotic scattering. We choose a plane $P$ of initial conditions in 
the energy shell, which intersects the important phase space objects like the stable and unstable manifolds of a NHIM and possibly the NHIM itself. This plane $P$ will become the domain of the indicator plot. 

Let us explain the numerical calculation of the indicator value 
$\Delta t$ for some point $x_0 \in P$. In the sense of scattering 
we assume that the initial point lies in the interaction region, and we 
also, choose an origin of the position coordinates lying in the 
interaction region.

We choose some value $\tau$ of the integration time and construct 
the trajectory through $x_0$ up to the time $\tau$ in the future and 
also up to the time $-\tau$ in the past by following the time-reversed 
dynamics. At the time $\tau$, the trajectory arrives at a final point $x^+$ at a
distance $d^+$ from the origin and with speed $v^+$. Then the time
delay for the future motion is $\Delta t^+ = \tau - d^+ / v^+$. Under the
time-reversed dynamics, the trajectory arrives at the time $-\tau$ at the
point $x^-$ at a distance $d^-$ from the origin and with speed
$v^-$. Accordingly, the time delay of the past motion is
$\Delta t^- = \tau - d^- / v^-$. 

Large values of the forward contribution $\Delta t^+$ indicate the stable 
manifolds and large values of the backward contribution $\Delta t^-$ 
indicate the unstable manifolds of localized invariant subsets. With 
increasing values of $\tau$, we detect higher levels of the fractal 
hierarchy of the infinity of tendrils of the invariant manifolds.
The sum of the forward and backward contributions gives the complete
indicator value $\Delta t$ defined as
\begin{equation}
\Delta t = \Delta t^+ + \Delta t^- = 2 \tau - d^+/v^+ - d^-/v^-
\end{equation} 
in analogy to Eq. 8 for the time delay of scattering.

%%%%%%%%

The time delay, as defined in Eqs. 8 and 9, becomes independent of the integration time $\tau$ if $\tau$ is so large that the initial point and the final point of the calculated trajectory segment have already reached the asymptotic region, i.e. left the interaction region. If the initial point and/or the final point still lie in the interaction region, then $d^+$ and/or $d^-$ remain small, and under an increase of $\tau$, the value of $\Delta t$ still increases. In such a case, the value of $\Delta t$ according to Eq. 9 is only a lower bound for the true asymptotic time delay. However, if the lower bound is already large, the true asymptotic time delay can only be larger. And in this sense, a large value of this lower bound is already a reliable indicator of the closeness to the stable and/or unstable manifolds of some invariant subset.

As a side remark, we give now a more detailed quantitative argument for this point. Since it is simpler and easier to understand for the scattering situation, we elaborate the arguments in the terminology of scattering and explain it first for the future time delay $\Delta t^+$.

Let us call $U$ the interaction region. We start a trajectory at time $t=0$ at some point inside of $U$, and we let the trajectory run until the time $t=\tau$. Now we distinguish two cases:

In the first case, at $t=\tau$ the trajectory has already left $U$, let us say it did so at a time $t_1 < \tau$. Then $d^+ / v^+$ is essentially $\tau - t_1$, in reallity it is a little larger since the distance run by the trajectory in the outgoing asymptotic region is a little smaller than $d^+$. Accordingly $\Delta t^+ = \tau - d^+ / v^+ $ is very close to $t_1$ ( in reallity a little smaller ) and it is a useful measure for the future time delay ( strictly speaking it is a lower bound ).

In the second case, at $t = \tau$ the trajectory did not yet leave $U$. Then the future time delay accumulated up to now is $\tau$. Remember that $d^+$ and $v^+$ are absolute values. Therefore $ - d^+ / v^+$ is always a negative value and $\Delta t^+ = \tau - d^+ / v^+  < \tau$. Accordingly $\Delta t^+$ is a lower bound to the time delay $\tau$ accumulated so far.

Finally, imagine that we increase $\tau$ from a value smaller than $t_1$ to a value larger than $t_1$. Then $\Delta t^+$ increases from the lower bound calculated in the second case to the approximately correct value calculated in the first case. This increase of $\Delta t^+$ is not necessarily monotonic in $\tau$. In any case, we obtain a lower bound to the true value. For the past time delay $\Delta t^-$ we apply exactly the same arguments under time reversal.

%%%%%%%

\begin{figure}[htbp]

(a)\includegraphics[scale=0.3]{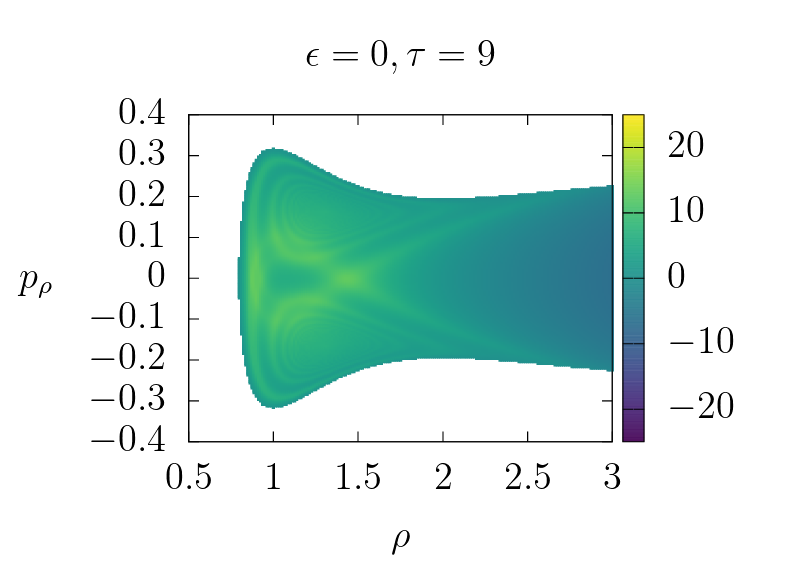}
(b)\includegraphics[scale=0.3]{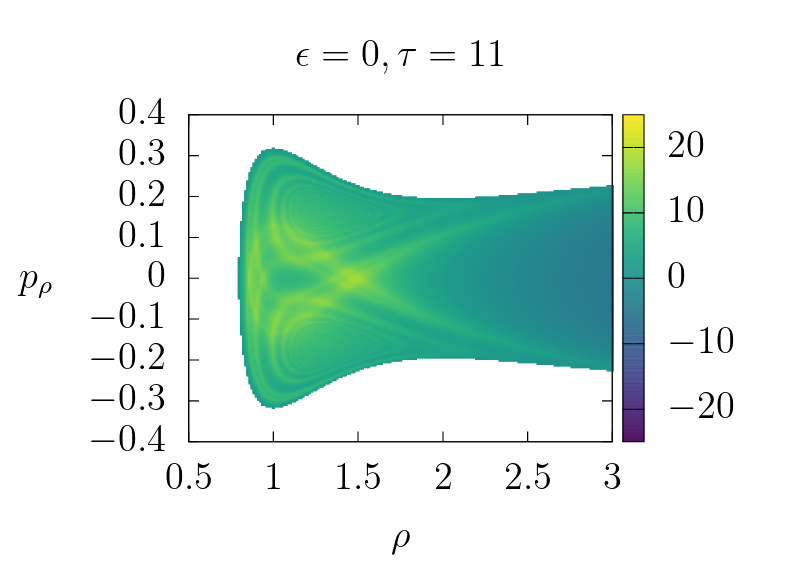}\\
(c)\includegraphics[scale=0.3]{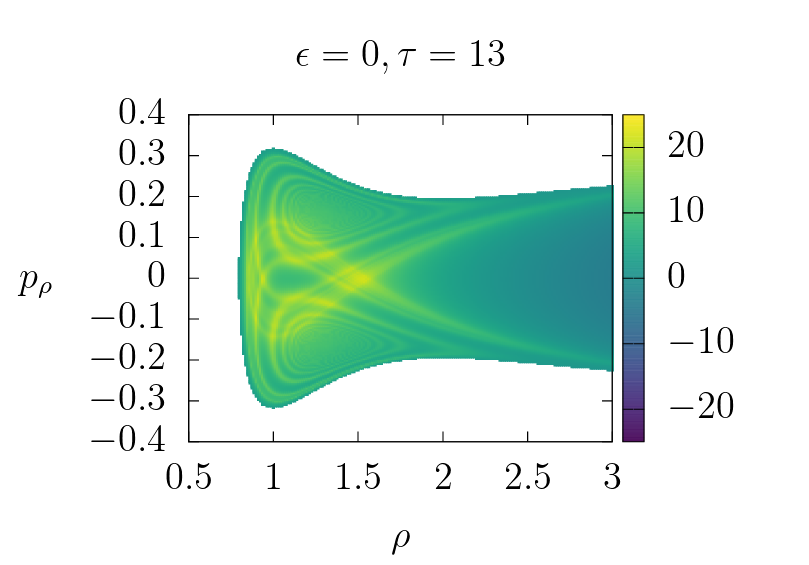}
(d)\includegraphics[scale=0.3]{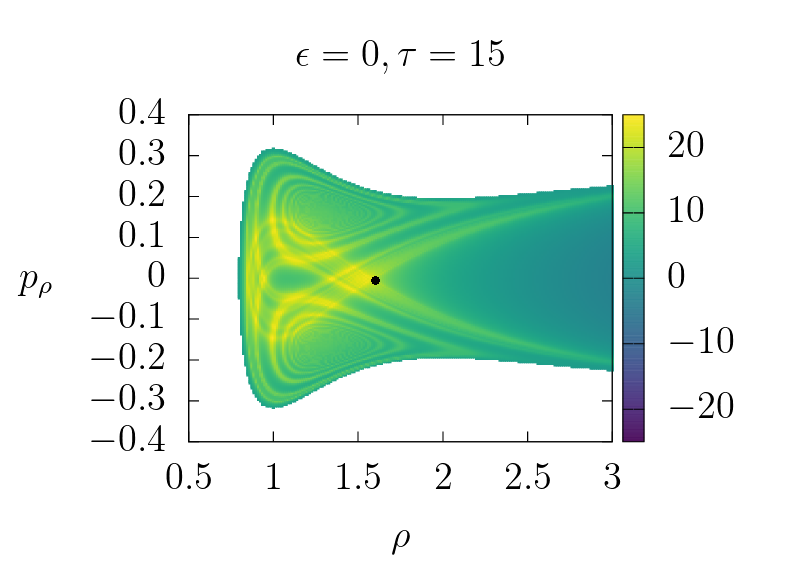}

\caption{ Delay time $\Delta t$ for the system symmetric under rotation around the $z$-axis, $L_z = 1$, and different integration times $\tau = 9,11,13,$ and $15$. The initial conditions are in the canonical plane $\rho$-$p_{\rho}$ with $L_z=1$ and $\phi$ arbitrary. We can see how the invariant manifolds of the NHIM $\mathcal{M}$ are revealed as the integration time $\tau$ is increased. The intersection between the set of initial conditions and the NHIM $\mathcal{M}$, the black point in the panel (d), is the hyperbolic fixed point $X_{L_z}$ for the Poincar\'e map of the reduced 2-dof symmetric system.}\label{fig:delay_time1}

\end{figure}

As an example to illustrate the use of the delay time as an indicator, first, let us consider the partially integrable system, i.e. when the electric charge is in the magnetic field of the dipole, and the perturbation is zero. The Hamiltonian of the system $H$ is given by Eq. 2 and initial conditions are chosen in the plane $p_\rho$-$\rho$, with $L_z=1$, and $z=0$ for an arbitrary angle $\phi$. The results for the delay time for the integration time $\tau = 9,11,13,$ and $15$ are displayed in Fig. \ref{fig:delay_time1}. The intersection of stable and unstable manifolds of the NHIM with the set of initial conditions is revealed as the value of $\tau$ grows. The intersections between the stable and unstable manifolds with the initial conditions are the points with the highest values of $\Delta t$ in the Fig. \ref{fig:delay_time1}(d). We can appreciate the complicated structure of the tangles between stable and unstable manifolds. In particular, in Fig. \ref{fig:delay_time1} (d) we can see clearly two symmetric spirals in the indicator plot for the symmetric 3-dof system. Also, the spirals are shown in the Poincar\'e maps in Fig. 2 of \cite{gj1} for different values of $L_z$. The intersection with the NHIM $\mathcal{M}$ is the black point $X_{L_z}$ in the Fig. \ref{fig:delay_time1}(d).

\newpage

\begin{figure}[htbp]

(a)\includegraphics[scale=0.3]{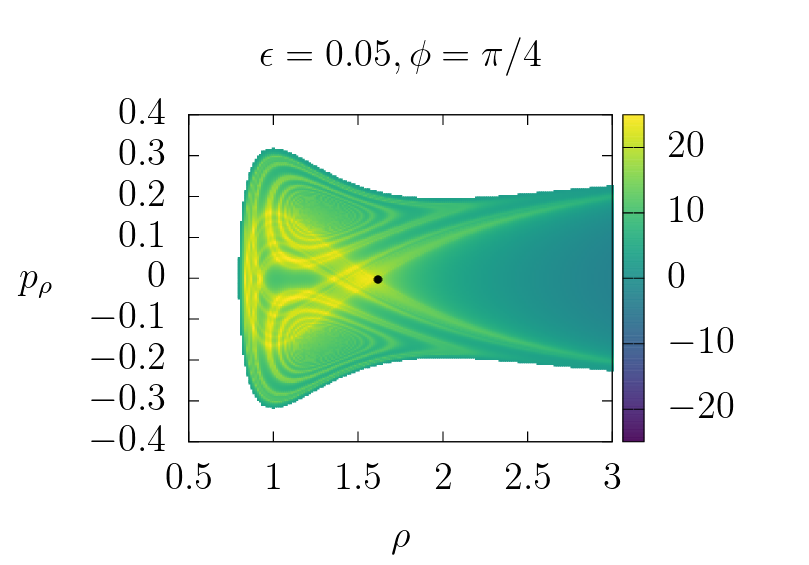}
(b)\includegraphics[scale=0.3]{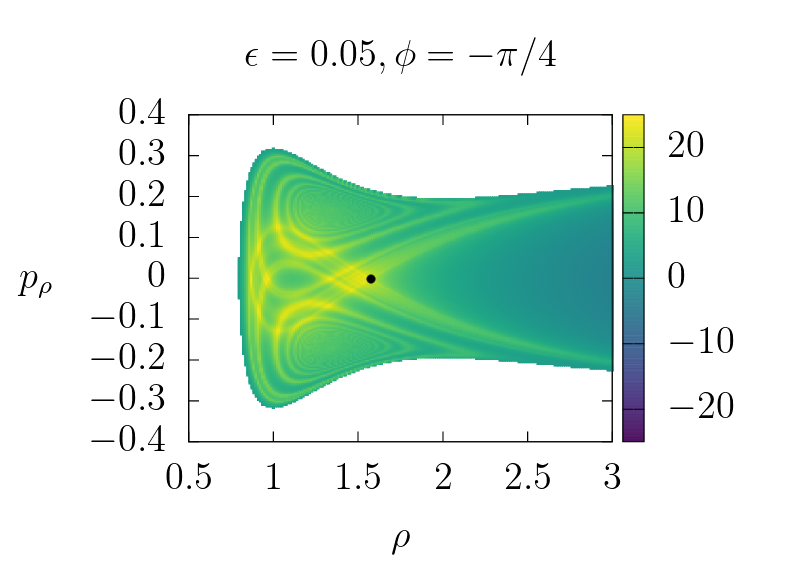}\\
(c)\includegraphics[scale=0.3]{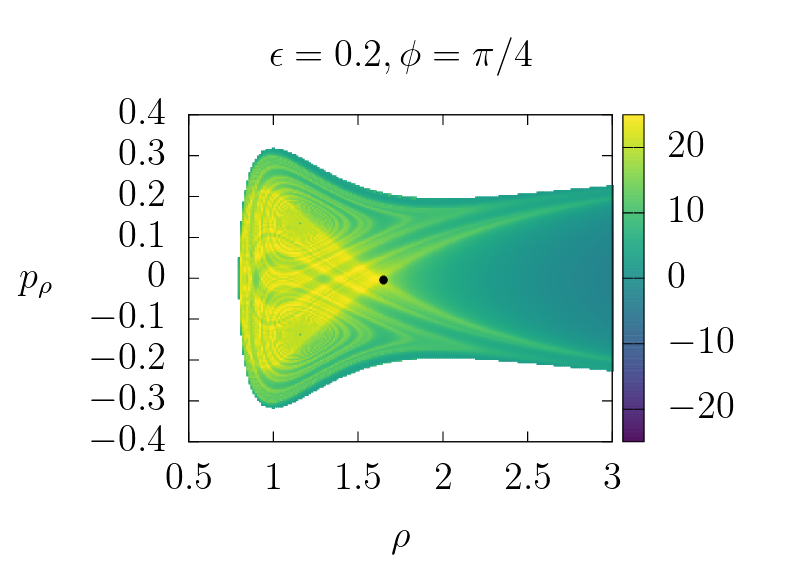}
(d)\includegraphics[scale=0.3]{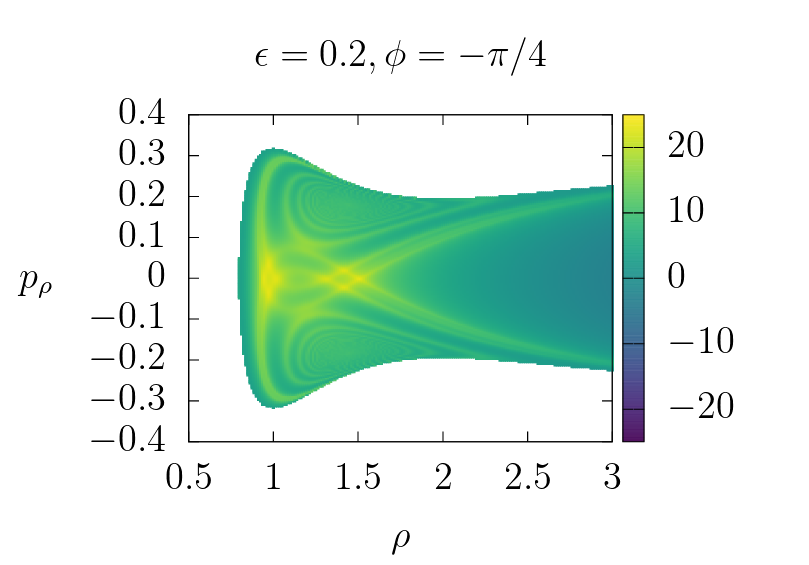}\\
(e)\includegraphics[scale=0.3]{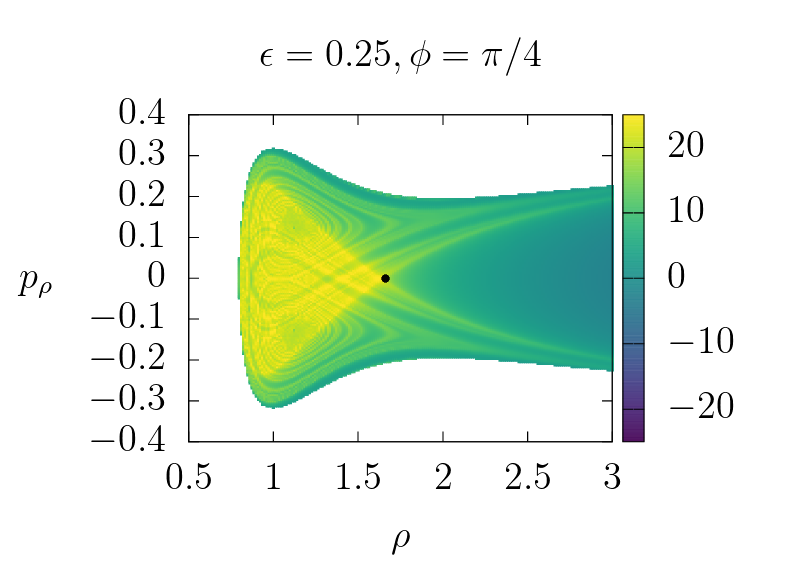}
(f)\includegraphics[scale=0.3]{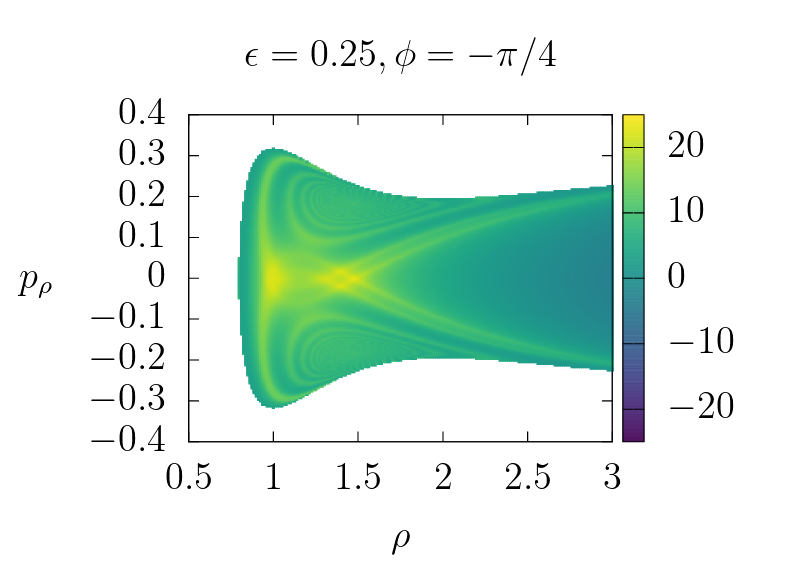}

\caption{ Delay time $\Delta t$ plotted over the $\rho - p_{\rho} $ plane at $L_z=1$ and $\phi = \pi/4$ in the panels (a), (c), and (e) or $\phi = -\pi /4$ in the panels (b), (d), and (f) respectively. The value of the perturbation parameter is $\epsilon = 0.05$ in the panels a) and (b), $\epsilon = 0.2$ in the panels (c) and (d) and $\epsilon = 0.25$ in the panels (e) and (f) respectively. The total energy still is $ E=0.05$ and the integration time is always $\tau = 15$. In the panels (a), (b), (c), and (e) the intersection point $x_h$ with the NHIM  $\mathcal{M}_\epsilon$ lies close to $\rho = 1.7$, $p_{\rho} = 0$, in the panels (d) and (f) this point is no longer a part of the NHIM  $\mathcal{M}_\epsilon$, however, it is still an accumulation point of homoclinic intersections between the stable and unstable manifolds and therefore still appears similar to an intersection of separatrix lines.
}\label{fig:delay_time2}

\end{figure}

Now, let us consider a perturbed system with an additional component in the magnetic vector potential corresponding to a magnetic quadrupole in the $z$-direction given by

\begin{equation}
A_z = \epsilon \frac{\rho^2 \sin{2\phi}}{(\rho^2+z^2)^{5/2}},
\end{equation}

\noindent where $\epsilon$ is the perturbation parameter. To see the changes in the phase space of the systems, let us calculate the delay time for different values of $\epsilon= 0.05,0.2,$ and $0.25$. The initial conditions are chosen in the same plane $p_\rho$-$\rho$, with $L_z=1$, $z=0$, for two angles $\phi=\pi/4,-\pi/4$. The results for the delay time are displayed in Fig. \ref{fig:delay_time2} for energy $E=0.05$. The left-hand side panels show the results for $\phi= \pi/4$ and on the right-hand side for $\phi=-\pi/4$. The regions where the indicator has a sharp maximum correspond to initial conditions on invariant manifolds. All the delay time calculations are done with the same integration time $\tau = 15$. 

In the parts (a), (b), (c), and (e) of the Fig. \ref{fig:delay_time2} we see two rather well-defined lines that separate regions of clearly different delay times. These two lines cross each other at a point $x_h$ at approximately $\rho = 1.7$, $p_{\rho} = 0$. The point $x_h$ is the intersection point of the NHIM $\mathcal{M}_\epsilon$ with the domain of the plot. And the two division lines crossing at $x_h$ are the intersections of the local segments of the stable and unstable manifolds  $W^{s/u}(\mathcal{M}_\epsilon)$ with the domain of the plot. As always, stable and unstable manifolds of codimension-2 invariant subsets act as separatrices between regions of distinct behaviour.

In contrast, in the parts (d) and (f) of Fig. \ref{fig:delay_time2} these divisions are already washed out. The reason is, that for $\epsilon = 0.2$ and $\epsilon = 0.25$ the remaining parts of the NHIM no longer intersect the $p_{\rho}$-$\rho $ plane for $\phi = - \pi / 4$. This decay of the NHIM $\mathcal{M}_\epsilon$ will be shown in more detail in the figures coming up in the next section. However, we know that the NHIM $\mathcal{M}_\epsilon$ itself is an accumulation set of homoclinic intersection surfaces. And when parts of the NHIM $\mathcal{M}_\epsilon$ survive the perturbation, then also parts of the homoclinic intersection surfaces survive. They are crossings of surviving parts of the stable and unstable manifolds $W^{s/u}(\mathcal{M}_\epsilon)$. And the corresponding bundles of segments of the stable or unstable manifold take over the role of the separatrix between different delay times also in regions of the phase space where the NHIM  $\mathcal{M}_\epsilon$ itself has disappeared. For this reason, we need a different method to study the decay of the NHIMs, which will be presented in the next section.

\section{The numerical procedure to follow a NHIM along a curve of the perturbation parameters}

In the previous section, we have seen the intersection of the stable manifold of the NHIM  $\mathcal{M}_\epsilon$ with some surface. In general, the trajectories close to this stable manifold $W^{s}(\mathcal{M}_\epsilon)$ can be very complicated. They can come close to the NHIM  $\mathcal{M}_\epsilon$ several times on very complicated trajectory segments. To obtain better information on the internal dynamics of $\mathcal{M}_\epsilon$ and represent its internal dynamics graphically, we can select very special intersection curves in the intersection plane, namely the intersections with the local segments of the stable manifold.

To understand the idea that we follow next, we have to remember the foliation theorem of the stable manifold, see \cite{wig}. Essentially it says that the stable manifold transports the internal dynamics of the NHIM along with it. And when a trajectory starts on some leaf of the stable manifold, it converges to the corresponding substructure of the NHIM and traces out this substructure as long as it is very close to the NHIM. When it starts close to the local segment of the stable manifold and is already close to the NHIM itself, it traces out the substructure over which it has started. And now, we describe a method to follow the NHIM and its internal dynamics through perturbations that utilizes this idea.

\begin{figure}[htbp]
\begin{center}
\includegraphics[scale=0.25]{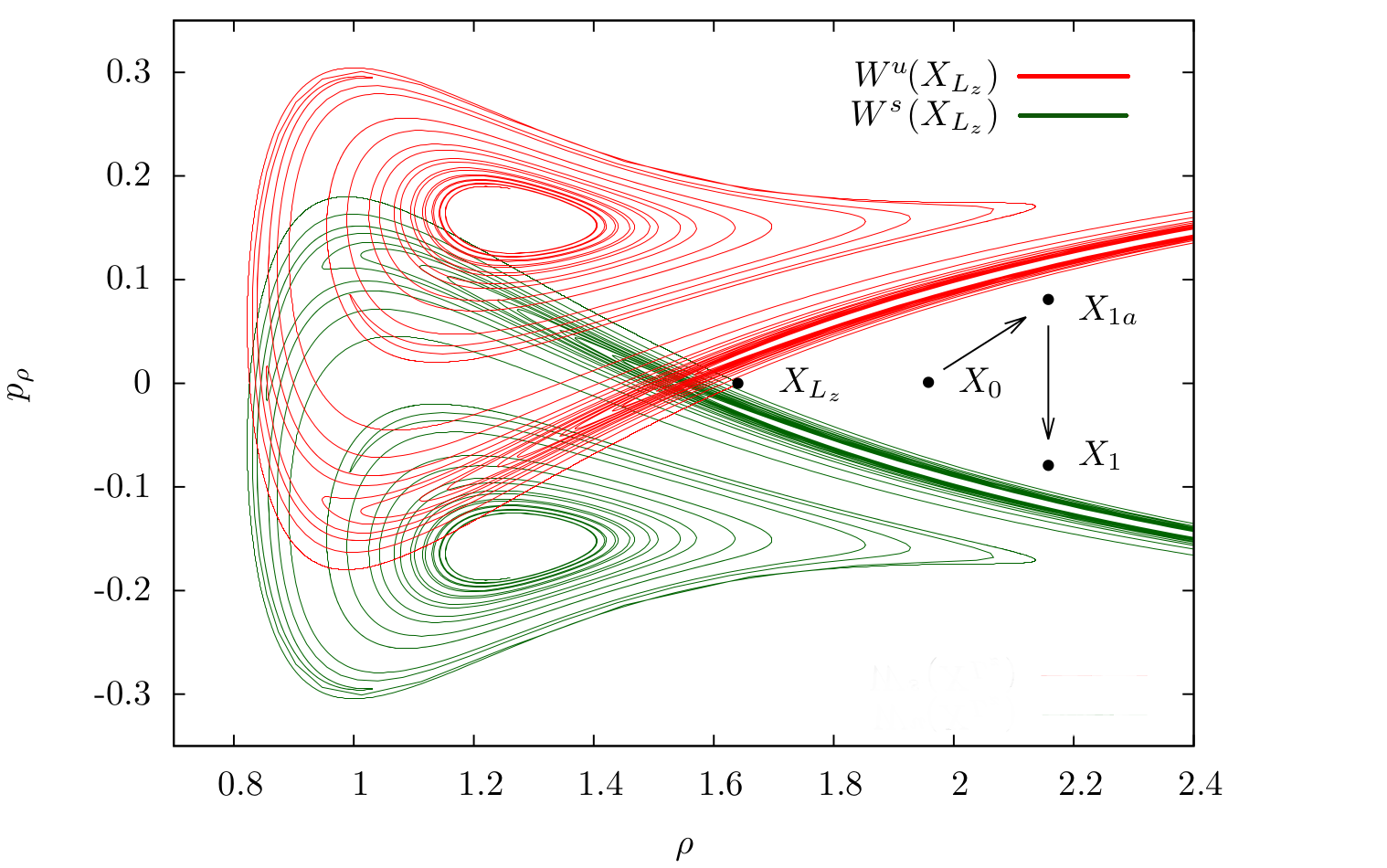}
\caption{ Algorithm to construct an aproximation to a NHIM and its internal dynamics. Let us consider the simplest example: the NHIM is the hyperbolic fixed point $X_{L_z}$ for a 2 dimensional Poincar\'e map. The stable and unstable manifolds $W^{s/u}(X_{L_z})$ are the green and red lines, repectively. The procedure to construct the NHIM has 3 basic steps: Find a point $X_{0}$ very close to $X_{L_z}$, calculate its image under the Poincar\'e map $X_{1a}$, and project the image of this point close to the stable manifold $W^{s}(X_{L_z})$ on the side where the dynamics is simple to get $X_{1}$. We can repeat this procedure to get the next iterations.}\label{fig:NHIM_construction}
\end{center}
\end{figure}

In order to illustrate the procedure to construct an approximation to a NHIM and its internal dynamics, let us consider a simple example: A hyperbolic fixed point of a 2-dimensional Poincar\'e map, see Fig. \ref{fig:NHIM_construction}. We want to obtain an approximation to the fixed point $X_{L_z}$ and its dynamics. First, we find a point $X_0$ close to the fixed point $X_{L_z}$ in the region where there are no segments of the homoclinic tangle between the stable and unstable manifolds $W^{s/u}(X_{L_z})$. In that region, the dynamics is simple, and the image of the point $X_0$, the point $X_{1a}$, is closer to the unstable manifold $W^{u}(X_{L_z})$. To avoid the next iteration going away from the hyperbolic fixed point $X_{L_z}$, we project the point $X_{1a}$ very close to the stable manifold $W^{s}(X_{L_z})$ in the same region where the dynamics is simple. In Fig. \ref{fig:NHIM_construction},  $X_1$ is the projection of the point $X_{1a}$. We take $X_1$ as the first stabilized iteration of the original point $X_0$ in this construction. To get the next point close to the NHIM $X_{L_z}$, we can repeat the previous procedure using the point $X_1$.

With the previous ideas in mind, let us consider a more general situation. Let us assume we have the NHIM given in the domain of the map for the parameter value $\epsilon_0$; let us call it $\mathcal{M}_0$. Next, we want to construct the NHIM numerically in the map domain for a parameter value $\epsilon_1$; let us call $\mathcal{M}_1$ the corresponding NHIM. We choose $\epsilon_1$ so close to $\epsilon_0$ that we are confident that $\mathcal{M}_1$ lies in a neighbourhood $U_1$ of $\mathcal{M}_0$. To construct a numerical approximation to a trajectory on $\mathcal{M}_1$, we apply the following procedure: We choose a line segment $A_{1,0}$ in $U_1$, e.g. the one with fixed coordinates $\rho, \phi, L$ and let $p_{\rho}$ run. The chosen segment should lie completely inside of $U_1$. We place a large number of points $a_j$ on $A_{1,0}$ and calculate the numerical trajectory for each $a_j$ and note for each one of them the time $t_j$ that this trajectory spends inside of $U_1$ until it leaves $U_1$ for the first time. Let us call this time the dwelling time of the trajectory in $U_1$. We look for the maximal value of $t_j$, let us say it is the one for $j=J$ and choose a subset $A_{1,1} \subset A_{1,0}$ which is the interval from $a_{J-1}$ to $a_{J+1}$. Then we place many points on $A_{1,1}$, calculate the dwelling time for all these points on $A_{1,1}$ and search again for the largest value and continue this zoom as long as our numerical accuracy allows. So we zoom in to the maximal value of the dwelling time along the curve segment $A_{1,0}$. The idea behind this search is: The stable manifold of the NHIM $\mathcal{M}_1$, call it $W^s(\mathcal{M}_1)$, has codimension 1 in the domain of the map. Then its local branch ( the part of this stable manifold that goes out directly from the NHIM before starting with its infinity of tendrils ) should intersect any reasonably chosen 1-dimensional curve in a point. And exactly at this point, the corresponding trajectory converges to the NHIM directly without leaving $U_1$ ever, and the dwelling time diverges. Of course, numerically, we only obtain an approximation to this point. However, we construct a numerical approximation of an initial point on $W^s(\mathcal{M}_1)$.

When we integrate numerically the trajectory belonging to this initial point, then it first approaches the NHIM $\mathcal{M}_1$, stays in the close neighbourhood of $\mathcal{M}_1$ for a finite time, and in the long run, leaves $\mathcal{M}_1$ again because of the combination of small errors in the initial condition and the normal instability of $\mathcal{M}_1$. 
 
Now let us assume we have found this intersection point as well as we can with our finite numerical precision. Then we let the trajectory run for a time that is clearly smaller than its dwelling time. We stop it at some intersection point with the domain of the map; let us say it is the point $p_f$. Then we take this final point as a new initial point, construct the line segment within $U_1$ going in $p_{\rho}$ direction through the point $p_f$ and take this new line segment as new $A_0$ and repeat the whole procedure described above. This gives a continuation of the trajectory segment constructed above. And so we continue until we have a large trajectory segment containing a lot of intersection points with the domain of the map. By this method, we obtain the corresponding trajectory in the map. The construction is a combination of the numerical trajectory with occasional projections on $W^s(\mathcal{M}_1)$. It can be interpreted as a version of the control of chaos \cite{con1, con2}, where we keep a numerical trajectory in the neighbourhood of an unstable invariant subset. The projections on $W^s(\mathcal{M}_1)$ counteract the numerical errors and the normal instability of the NHIM $\mathcal{M}_1$. Remember that we always project onto the local segment of the stable manifold, and this is why the obtained trajectory traces out the substructure over which it has started.

If we have a sufficient number of points of a trajectory, then we choose a new independent initial point and construct the corresponding numerical approximation of the trajectory on the NHIM $\mathcal{M}_1$, etc. After having done this construction for many different initial points distributed over the whole $\mathcal{M}_1$, we have many points very close to $\mathcal{M}_1$ and also have many trajectories on the Poincar\'e map restricted to $\mathcal{M}_1$. This gives first the surface of the NHIM $\mathcal{M}_1$ in the domain of the map. And second, it gives a graphical representation of the dynamics restricted to the NHIM $\mathcal{M}_1$. It is a representation of the 2-dof internal dynamics of the NHIM.

Next, we choose a new value $\epsilon_2$ of the perturbation parameter, choose as $U_2$ an appropriate neighbourhood of $\mathrm{NHIM}_1$ and repeat the whole method to obtain $\mathrm{NHIM}_2$, etc. This provides a sequence of NHIM constructions along a path in the parameter space. The corresponding plots of the restricted map give a pictorial representation of the development scenario of the NHIM. For an application of this search strategy to the dipole problem, see \cite{gj2},  for the application to several problems from celestial mechanics, see \cite{jz1, zj2, zj3, zj4} and for the application to hydrogen in a rotating field, see \cite{jw}.

In this procedure, we construct the intersection between a line of initial points and the local branch of the stable manifold of the NHIM by searching for the point with the maximal dwelling time in an appropriate neighbourhood. There is an alternative way to find this intersection point for some examples (this includes the magnetic dipole problem). Let us mention this alternative method very briefly for the example of the dipole. We know that the NHIM has the shape of a cylinder segment, which divides the domain at least locally into an interior and an exterior part. On the part of the line of initial conditions on one side of the true intersection point, the trajectories run to the interior and on the other side of the line of initial conditions, they run to the exterior. We define the neighbourhood $U$ as a layer around the cylindrical NHIM, and this layer has two boundaries, an inner one and an outer one. As line segment $A$, we choose a straight line segment that connects the inner and the outer boundary. And the intersection of this line segment $A$ with the stable manifold of the NHIM is the division point between crossing first the inner and crossing first the outer boundary. Then we can search for this division point with a bisection method. As long as it works reliably, this bisection method is numerically faster.

\begin{figure}[htbp]
\begin{center}
	\includegraphics[scale=0.5]{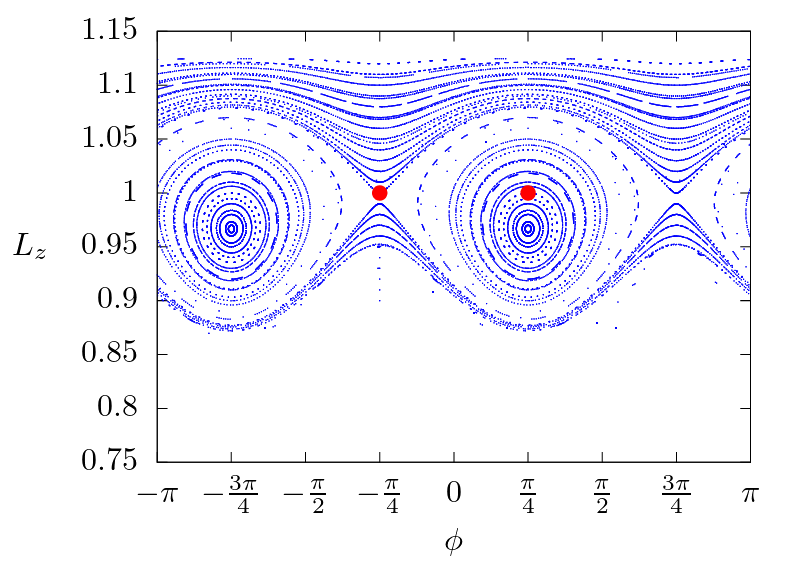}
	\caption{ Poincar\'e map of the internal dynamics of the NHIM  $\mathcal{M}_\epsilon$ for $\epsilon = 0.05$. The perturbation breaks the rotational symmetry around the $z$-axis and 2 big KAM islands are formed. The red dots correspond to the initial conditions for the corresponding plots of $\Delta t$ in Figs. 4(a) and 4(b).  }
	\label{fig:Poincare_map_005}
\end{center}
\end{figure}

\begin{figure}[htbp]
\begin{center}
	\includegraphics[scale=0.5]{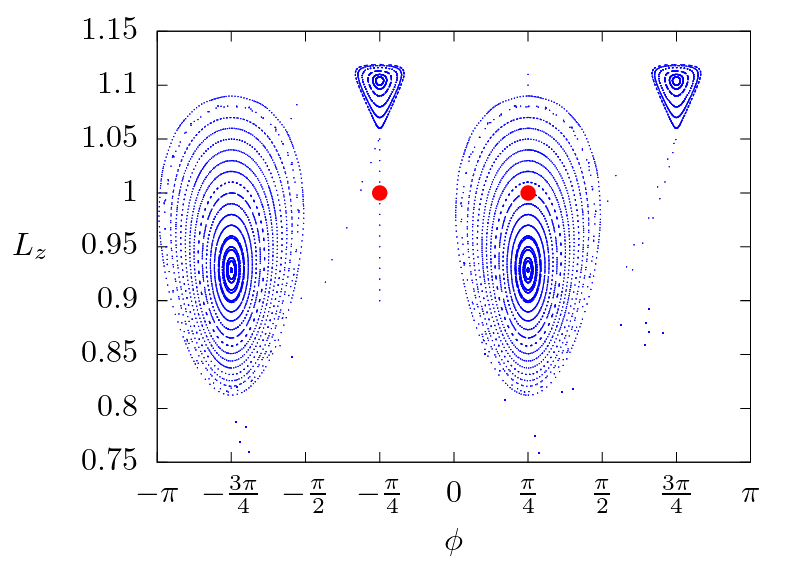}
	\caption{ Poincar\'e map of the internal dynamics  of the NHIM $\mathcal{M}_\epsilon$ for $\epsilon = 0.2$. The perturbation has already destroyed the primary KAM curves outside of two main KAM islands, and two KAM islands with centres at $\phi=-\pi/4, 3\pi/4$ grow out of the upper bounday. The red dots correspond to the initial conditions for the corresponding plots of $\Delta t$ in Figs. 4(c) and 4(d). }
	\label{fig:Poincare_map_02}
\end{center}
\end{figure}

\begin{figure}[htbp]
\begin{center}
	\includegraphics[scale=0.5]{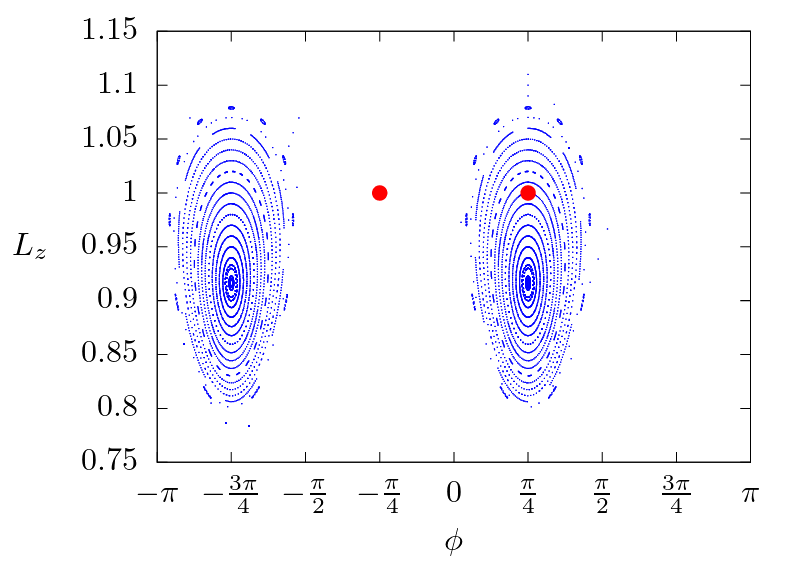}
	\caption{ Poincar\'e map of the internal dynamics of the NHIM $\mathcal{M}_\epsilon$ for $\epsilon = 0.25$. The perturbation destroyed the two upper KAM islands and only the two main KAM islands survive. The red dots correspond to the initial conditions for the corresponding plots of $\Delta t$ in Figs. 4(e) and 4(f).  }
	\label{fig:Poincare_map_025}
\end{center}
\end{figure}

Let us consider the NHIM $\mathcal{M}_\epsilon$ in the dipole system calculated with the procedure described before. As in the previous section we choose again the values $\epsilon = 0.05, 0.2, 0.25$ for the perturbation parameter and the energy is still $E=0.05$. To visualize the NHIM $\mathcal{M}_\epsilon$, it is convenient to take the Poincar\'e surface of section $z=0$ with both orientations because of the discrete reflection symmetry in $z$. To visualize the iterations of the Poncar\'e map,  we take its projection onto the plane $L_z$-$\phi$, like in \cite{gj2}. Figs. \ref{fig:Poincare_map_005},\ref{fig:Poincare_map_02},\ref{fig:Poincare_map_025} show the results. We can appreciate how the rotational symmetry around the $z$-axis is broken, and the NHIM bifurcates.

We observe the following important events in the development scenario of the internal dynamics of the NHIM. For small perturbations, as in Fig. 6, many primary KAM curves survive. Only the primary curve with winding number 1/2 causes large scale secondary structures already for very small values of $\epsilon$. From this primary KAM curve, a pair of stable points of period 2 survives at $\phi = -3 \pi /4$ and $\phi = \pi /4$. These two stable points become the centres of two large secondary KAM islands. In addition a pair of unstable points of period 2 survives at $\phi = -\pi /4$ and $\phi = 3 \pi /4$. These unstable points become the organizing centres of a fine chaos strip which looks like a separatrix curve between the large secondary KAM islands and the surviving primary KAM lines. The red dots in the figures are the points that correspond to the
values of $\phi$ and $L_z$ for the domains of the indicator plots in Fig. 4.

For increasing perturbation values, the primary KAM lines disappear quite rapidly. And for $\epsilon = 0.2$, remnants of the NHIM are essentially 4 KAM islands, see Fig. 7. Two of them are the large KAM islands of period 2 with centres at $\phi = -3 \pi /4$  and $\phi = \pi /4$ already found for very small values of $\epsilon$. The other two islands are also of period 2, have centres at $\phi = - \pi /4$ and  $\phi = 3 \pi /4$, and they have grown out of the upper boundary of the domain of the map. After the destruction of the primary KAM lines, there are no boundaries on the former NHIM surface, which prevent general trajectories outside of the secondary KAM islands from moving to the former NHIM surface's lower boundary and disappearing.

When we increase the value of $\epsilon$ further, then the new islands coming out of the upper boundary move down to lower values of $L_z$, and their centre points of period 2 collide at $\epsilon \approx 0.228 $ with the unstable points of period 2 coming from the primary KAM line of winding number 1/2. In the collision, these two pairs of points of period 2 disappear in a saddle-centre bifurcation. As seen numerically in Fig.8, afterwards, only the two large period 2 KAM islands are the last remnants of the NHIM. In particular, observe that in  Figs. 7 and 8, the red dot at $\phi = -\pi /4$ no longer belongs to the surviving NHIM surface. Accordingly, the NHIM no longer intersect the domains of the indicator plots in Figs. 4(d) and (f). This explains the blurred structures already seen in Figs. 4(d) and 4(f).

Fig. 9 shows the values of the tangential and the normal eigenvalues of the hyperbolic points of period 2. The parallel eigenvalue $\lambda_\parallel$ is always close to 1 and returns exactly to 1 in the saddle-centre bifurcation at $\epsilon \approx 0.228$, where the curve ends. In contrast, the normal eigenvalue $\lambda_\perp$ is always larger, which shows that this hyperbolic point always belongs to the NHIM surface as long as it exists.

\begin{figure}[htbp]
\begin{center}
	\includegraphics[scale=0.25]{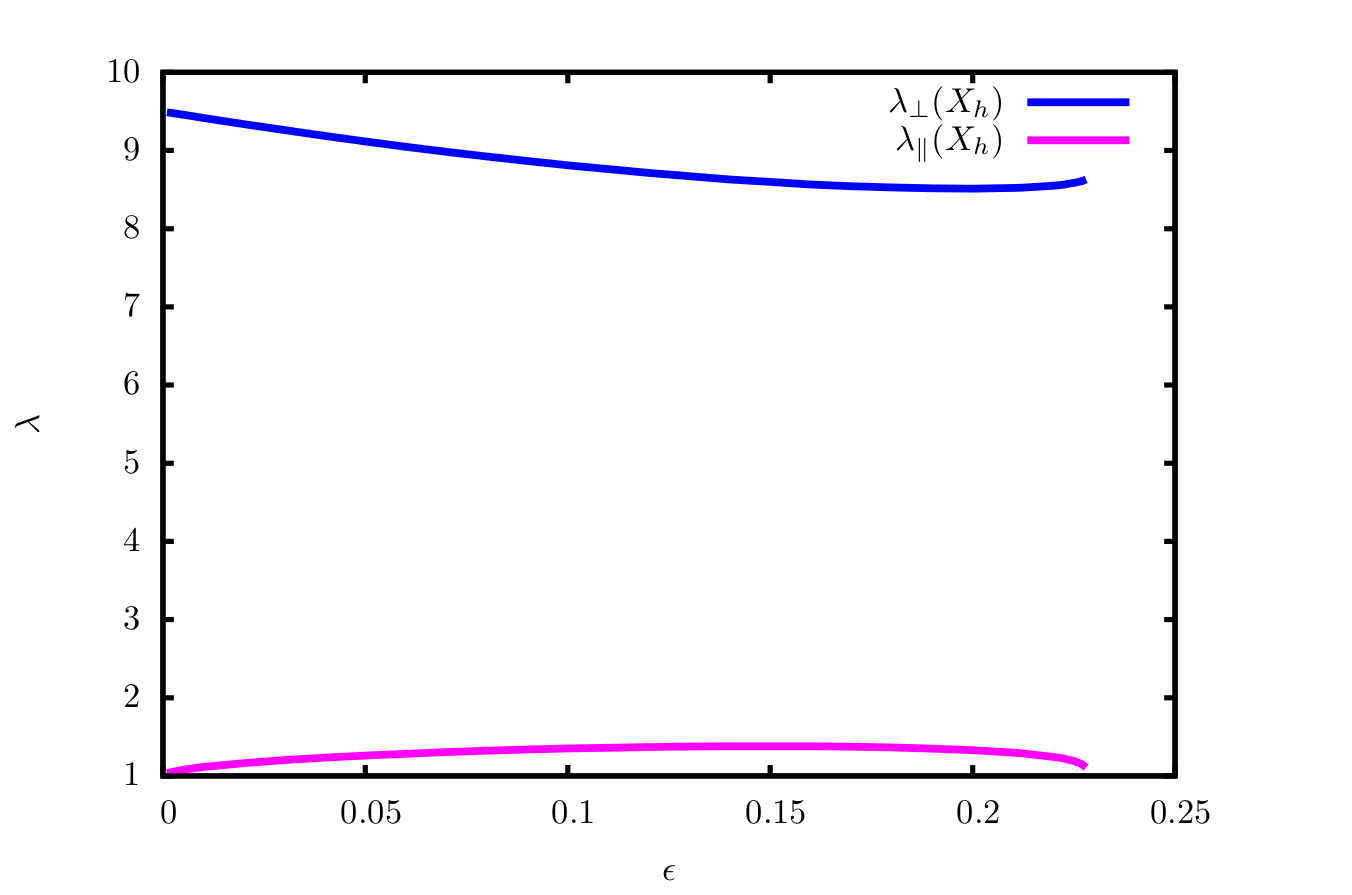}
	\caption{ Eigenvalues of the hyperbolic period 2 point $X_h$ in the line $\phi = -\pi/4$ in the Poincar\'e map of the NHIM $\mathcal{M}_\epsilon$ in Figs. \ref{fig:Poincare_map_005},\ref{fig:Poincare_map_02},\ref{fig:Poincare_map_025}. The eigenvalues $\lambda_\parallel$ and $\lambda_\perp$ correspond to the eigenvalues on the parallel and normal directions to the $\mathcal{M}_\epsilon$ respectively.}
	\label{fig:eingenvalues}
\end{center}
\end{figure}

\newpage

\section{Final remarks}

We have discussed methods to identify the points sitting on stable manifolds of unstable invariant subsets and corresponding methods to search numerically for NHIMs. We have put these ideas into a larger frame, pointing out their relations. The basic idea is always based on the following observation: If the initial point of a trajectory sits exactly on the stable manifold of an invariant subset, then the trajectory will converge to this invariant subset and the time delay becomes infinite. Numerically, we observe a large time delay in some subset of the phase space. This idea has also been employed before in the investigation of chaotic scattering, where general scattering functions and, in particular, the time delay in the interaction region become singular for asymptotic initial conditions on the stable manifold of the chaotic saddle. In this sense, scattering functions have always been used as indicators on the set of incoming asymptotes.

Figures \ref{fig:delay_time1} and \ref{fig:delay_time2} of the present article demonstrate this idea graphically by plotting the time delay in the inner region of the phase space over a coordinate plane of initial conditions. We see clearly the lines representing the intersections of the plane of initial conditions with the stable and unstable manifolds of the NHIM, and we also recognize crossings of the stable and unstable manifolds. However, with the information provided by Fig. \ref{fig:delay_time2} alone, it was not yet possible to distinguish between crossings representing the NHIM itself and crossings representing general homoclinic intersection surfaces. Therefore, we had to go one step further and identify the intersections of the plane of initial conditions with the local segments of the stable manifold. The corresponding filter criterion is that the trajectory converges to the direct neighbourhood of the NHIM monotonically without making large excursions. And this filter was also essential to construct the plots of the internal dynamics of the NHIM, as shown in Figs. \ref{fig:Poincare_map_005},\ref{fig:Poincare_map_02},\ref{fig:Poincare_map_025} and explained in section 4.

We have seen that the internal dynamics is dominated by rather regular structures, at least in such parts of the NHIM that have survived the actual perturbation. Now, the reader may ask whether or not it is general that the internal structure of a NHIM is mainly regular or whether this is an exceptional property of the dipole system. The numerical results of many different systems indicate that it seems to be general; see again the above mentioned references \cite{zj2, zj3, zj4, jw}. The following arguments may give some explanation. We start to investigate the development scenario of the NHIM in one of the simple cases explained in section 2. In particular, we take the simple case for which the internal dynamics of the NHIM is exactly integrable.

The persistence theorem guarantees that the NHIM persists when the tangential instability is smaller than the normal instability. It does not claim that the NHIM must decay in regions where the tangential instability becomes larger than the normal instability. However, there are some indications that this decay is the usual event. First, in \cite{fen} it is mentioned that the NHIM surface grows spikes as soon as the tangential instability becomes larger than the normal one. This phenomenon has also been observed in a model map \cite{dgj}. Here the tangential instability of a fixed point in the NHIM is parameter dependent, and as soon as the local tangential instability becomes larger than the normal instability, the fixed point becomes the tip of a spike and leaves the NHIM. Also, example 1.1 in section 1.2.1 of \cite{je} suggests the formation of singularities in the NHIM as soon as the tangential instability becomes equal to the normal instability. This behaviour seems to be common.

The survival of only regular parts of the NHIM is exactly what we see graphically in the plots of the development scenario in the Figs.\ref{fig:Poincare_map_005},\ref{fig:Poincare_map_02},\ref{fig:Poincare_map_025}. In addition, this closeness to internal integrability implies that the homoclinic trajectories connect a particular initial substructure of the NHIM with some particular final substructure. For more information on the decay of NHIMs and the related transient effects, see also \cite{c21}. This closeness to the integrability of the internal dynamics makes evident why we consider it important to apply strategies for the search of NHIMs that automatically also provide a pictorial display of the internal dynamics of the NHIMs. 

There are other different methods to calculate NHIMs using the normal form techniques and the parametrization method; see references \cite{wsw,h,ch1,ch2}. Those methods are able to follow perturbed NHIMs that persist. The method described in the present article is an alternative to study the bifurcations and decay of NHIMs because it is based on trajectories and does not require any extra hypothesis.

A new side result in section 2 is the idea of how the effective potential of a reduced system can have index-1 saddles carrying invariant subsets, even if the effective potential of the full-dimensional system does not have corresponding nondegenerate saddles. Here the magnetic dipole problem is a very drastic example since the effective potential of the full-dimensional system is identically zero.

The basic ideas presented in this article also work for even more degrees of freedom. The only complication for 4 or more degrees of freedom is the graphical presentation of the internal dynamics of the codimension-2 NHIMs. For $n$-dof systems, the NHIM surface in the domain of the map is $2n-4$ dimensional, and for $n>3$, it can no longer be displayed as a 2-dimensional plot.

\section*{Acknowledgment}

C J thanks DGAPA-UNAM for financial support under grant number IG101122 and CONACyT for financial support under grant number 425854. F G M thanks CIC AC UNAM.

\end{document}